\documentclass[sigconf]{acmart}

\usepackage{enumitem}
\setlist[itemize]{leftmargin=*}
\usepackage{multirow}
\usepackage{makecell}
\usepackage{amstext}
\usepackage{geometry}
\AtBeginDocument{%
  \providecommand\BibTeX{{%
    \normalfont B\kern-0.5em{\scshape i\kern-0.25em b}\kern-0.8em\TeX}}}

\copyrightyear{2022}
\acmYear{2022}
\setcopyright{acmcopyright}\acmConference[CIKM '22]{Proceedings of the 31st ACM International Conference on Information and Knowledge Management}{October 17--21, 2022}{Atlanta, GA, USA}
\acmBooktitle{Proceedings of the 31st ACM International Conference on Information and Knowledge Management (CIKM '22), October 17--21, 2022, Atlanta, GA, USA}
\acmPrice{15.00}
\acmDOI{10.1145/3511808.3557544}
\acmISBN{978-1-4503-9236-5/22/10}

\begin{document}
\begin{sloppypar}
\title{AMinerGNN: Heterogeneous Graph Neural Network for paper click-through rate prediction with fusion query}

\author{Zepeng Huai$^{\dagger}$$^{\ddagger}$}
%\authornote{Work performed during the internship at AMiner.}
\authornote{Corresponding Author.}
\email{zepenghuai6@gmail.com}
\orcid{0000-0003-3741-5157}
\affiliation{%
  \institution{$^{\dagger}$School of Artificial Intelligence, UCAS}
  \institution{$^{\ddagger}$CASIA}
  \city{Beijing}
  \country{China}
}

\author{Zhe Wang}
\email{zhe.wangz@bytedance.com}
\affiliation{%
	\institution{ByteDance Inc.}
	\city{Mountain View, CA}
	\country{United States}
}

\author{Yifan Zhu}
\email{zhuyifan@tsinghua.edu.cn	}
\affiliation{%
	\institution{Tsinghua University}
  \city{Beijing}
\country{China}
}

\author{Peng Zhang}
\email{peng.zhang@aminer.cn}
\affiliation{%
	\institution{Zhipu AI Lab}
	\city{Beijing}
	\country{China}
}

\begin{abstract}
	Paper recommendation with user-generated keyword is to suggest papers that simultaneously meet user's interests and are relevant to the input keyword. This is a recommendation task with two queries, a.k.a. user ID and keyword. However, existing methods focus on recommendation according to one query, a.k.a. user ID, and are not applicable to solving this problem. In this paper, we propose a novel click-through rate (CTR) prediction model with heterogeneous graph neural network, called AMinerGNN, to recommend papers with two queries. Specifically, AMinerGNN constructs a heterogeneous graph to project user, paper, and keyword into the same embedding space by graph representation learning. To process two queries, a novel query attentive fusion layer is designed to recognize their importances dynamically and then fuse them as one query to build a unified and end-to-end recommender system.  Experimental results on our proposed dataset and online A/B tests prove the superiority of AMinerGNN. 
%	To our best knowledge, this is the first attempt to study recommendations with two queries.
\end{abstract}



\maketitle

\section{Introduction} \label{sec:introduction}
\begin{figure}[t]
	\centering
	\includegraphics[width=\linewidth]{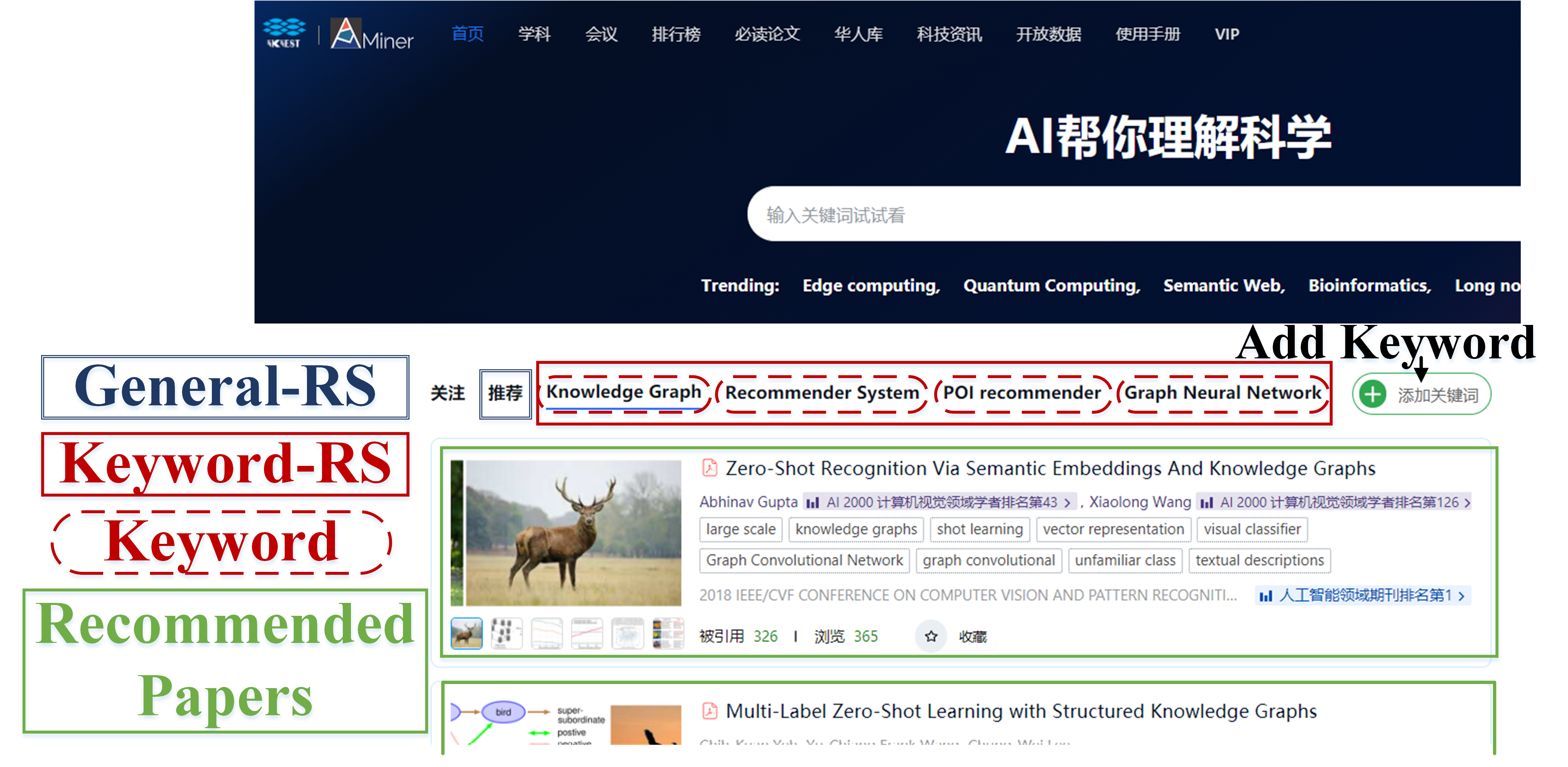}
	\caption{Keyword-RS in AMiner. Best viewed in color.}
	\Description{Keyword-RS in AMiner.}
	\label{fig:Keyword-RS}
\end{figure}

AMiner\footnote{www.aminer.com} is an academic information retrieval website, which can provide paper recommendation to researchers. Depending on whether the user inputs a keyword or not, there are two recommender systems (RSs) for online services, called General-RS and Keyword-RS (as shown in Figure \ref{fig:Keyword-RS}). When simultaneously given the user ID and a user-generated keyword (usually a research direction, such as \textit{graph neural network} or \textit{pretraining}), Keyword-RS suggests papers that not only are relevant to this keyword but also satisfy user potential interests.

According to the relation between the user-generated keyword and the research direction of the user based on his interacted papers, there are three segmentation scenarios in Keyword-RS: (1) \textbf{both of them refer to the same research field (S1)}; (2) \textbf{there exists the potential of interdisciplinary research between them (S2)}; (3) \textbf{they are independent and have no relevance (S3)}. We collect some user feedbacks and have an important finding: \textbf{the purpose that researchers use AMiner and the importance of these two queries vary according to scenarios}, which are as follows: (1) In \textbf{S1}, users aim to make a deeper literature review about this research field, like tracking the advanced methods or seeking the highly cited papers. At this time, Keyword-RS should utilize his interests to recommend and user ID is more important than keyword since user ID contains the personal information. (2) In \textbf{S2}, users intend to know how this approach (user-generated keyword) can be integrated into his past research, like how contrastive learning can be used for recommendation. Two queries are both useful in this situation.  (3) In \textbf{S3}, users prefer to improve the general understanding of this new field (we call it a new field since this keyword never appears in his interacted papers), like Bert \cite{devlin2018bert} in the field of Natural Language Processing. Therefore, it is reasonable to recommend papers without personal preferences and show some typical or hot papers clicked by most researchers, which means the keyword is more important than user ID. To satisfy the above recommendation requirements, there are two problems: (1) \textbf{how to model the relation between the user-generated keyword and his past research direction}; (2) \textbf{how to build a unified and end-to-end recommender system applicable in all scenarios}. 

The above business is a typical click-through rate (CTR) prediction task but with a novel input, a.k.a.  user ID and keyword. 
Early works predict CTR via automatic feature interactions like DeepFM \cite{guo2017deepfm}, PNN \cite{qu2016product} and InterHAt \cite{li2020interpretable}. 
Recent researchers aims to model user behaviors (e.g. DIN \cite{zhou2018deep}, UBR \cite{qin2020user} and SIM \cite{pi2020search}) and time-varying interests (e.g. DIEN \cite{zhou2019deep} and SURGE \cite{chang2021sequential}). 
However, all of them are given just one query, a.k.a. user ID, and are less applicable. 
Another close solution is the methods in product search by regarding the input keyword as the search query. 
They focus on inferring personalized search inclination (e.g. HEM \cite{ai2017learning} and GEPS\cite{zhang2019neural} ) or analysing users’ long and short-term preferences (e.g. ALSTP \cite{guo2019attentive}  and MGDSPR \cite{li2021embedding}).
However, they neglect that the importances of personalization and search-aware intention are dynamically changing with scenarios.

To address the foregoing problems, we propose an end-to-end heterogeneous graph neural network for paper click-through rate (CTR) prediction, called AMinerGNN. It consists of two components: (1) \textbf{Graph embedding layer}.  To model the semantic correlation in Keyword-RS, we construct a heterogeneous graph, which consists of user, paper and keyword,  and project the above three types of entities into the same embedding space via graph based representation learning, which is the basic to further learn the relation between user-generated keyword and his past research direction. (2) \textbf{Query attentive fusion layer}. We employ an attention unit to automatically mine user scenario-specific purpose and dynamically adjust the importances of two queries, which are further combined as one query to build a unified recommender system. 

\section{RELATED WORKS}
We briefly summarize two related subareas of information retrieval systems, which are click-through rate prediction and product search, as follows. 

Click-through rate (CTR) estimation plays as a core function module in various personalized online services. Feature interaction \cite{rendle2010factorization,juan2016field,qu2016product,guo2018deepfm} is an early method to build a predictive model. FM \cite{rendle2010factorization} assigns a k-dimensional learnable embedding vector to each feature and explores their 2-order interactions. DeepFM \cite{guo2018deepfm} incorporates the deep network to learn high-order feature interactions effectively.
In recent years, modeling user behaviors \cite{zhou2018deep,zhou2019deep,qin2020user,pi2020search,chang2021sequential} is becoming an essential topic since they contain crucial patterns of user interests. DIN \cite{zhou2018deep} introduces an attention mechanism to attribute different historical behaviors with different scores according to the relatedness with the target item. Based on DIN, DIEN \cite{zhou2019deep} utilizes GRU structure to capture the temporal interests. UBR \cite{qin2020user} proposes a new framework that uses search
engine techniques to retrieve the most relevant behaviors, which not only solves the time complexity
problem but also alleviates the noisy signals in long consecutive sequences.
SURGE \cite{chang2021sequential} performs cluster-aware and query-aware graph convolutional propagation to extract users' current activated core interests from long but noisy behavior sequences.
However, their inputs are userID and target item, which can not directly resolve the CTR problem with fusion query.

Product search is an important way for people to browse and purchase items on E-commerce platforms. 
A basic category of methods \cite{duan2013probabilistic,duan2015mining,duan2013supporting,karmaker2017application} is to associate the free-form user queries with structured data stored in relational databases. MAM \cite{duan2013probabilistic} proposes a probabilistic retrieval model to mine useful knowledge from product search log data.  \cite{karmaker2017application} discusses issues and strategies when
introducing learning-to-rank (LETOR) methods to the product search. 
Recent works aim to study the user’s personal and temporal preferences \cite{ai2017learning,guo2019attentive,zhang2019neural,bi2020transformer,li2021embedding}. HEM \cite{ai2017learning} attentively combines userID and query to predict personal behavior by mapping the user into the same latent space with the query and product. ALSTP \cite{guo2019attentive} integrates the long and short-term user interests with the current query for the personalized product search. MGDSPR \cite{li2021embedding} simultaneously models query semantics and historical behaviors, aiming at retrieving more products with good relevance.
However, all of them are under-explored in the task where the search query and personal preferences have non-uniform relations.

\section{Preliminaries}
In this section, we first formulate the problem and then give a detailed description of the construction of the heterogeneous graph used in Keyword-RS.

\subsection{Problem Setup}
\textbf{Paper CTR with two queries}. Given a set $<\mathcal{U}, \mathcal{K}, \mathcal{P}>$, where $\mathcal{U}=\{u_1, \cdots, u_M\}$ denotes $M$ users, $\mathcal{K}=\{k_1, \cdots, k_T\}$ denotes $T$ keywords and $\mathcal{P}=\{p_1, \cdots, p_N\}$ denotes $N$ papers. In this problem, user and keyword are both regarded as queries, leading to three kinds of interaction data, which are user-paper, keyword-paper and user\&keyword-paper interactions. The above three interactions are represented as $\mathcal{Y}=\{y_{u-p},y_{k-p},y_{uk-p}|u \in \mathcal{U}, k \in \mathcal{K}, p \in \mathcal{P}\}$, respectively. $y_{uk-p}$ indicates whether $u$ clicks $p$ when $u$ inputs keyword $k$. $y_{u-p}$ means whether $u$ clicks $p$ in Keyword-RS or searches $p$ in AMiner. $y_{k-p}$ represents whether $p$ is clicked when the input keyword is $k$. Note that $y_{u-p}$ leverages both search and recommendation log while the other two just use recommendation log. We also utilize two widely used information \cite{zhang2021deep, li2021leveraging,liu2019feature} to improve CTR performance: (1) query behaviors including two aspects: user behaviors $H_u=\{p| y_{u-p}=1\}$ and keyword behaviors $H_k=\{p| y_{k-p}=1\}$, (2) item features $F^{p}=[F^{p}_1, \cdots, F^{p}_q, \cdots]$, where $F^{p}_q$ denotes $q$-th feature of paper $p$ like citation number or publication year. Note that $H_u$ reckons without keywords and represents papers that $u$ clicked, while  $H_k$ denotes papers all users clicked when the input keyword is $k$. When user $u$ inputs a keyword $k$, the Keyword-RS is to prediect whether $u$ will click $p$ as $\hat{y}_{uk-p}=f(u,k,p,H_u,H_k,F_p)$.
%\begin{equation}
%	\hat{y}_{uk-p}=f(u,k,p,H_u,H_k,F_p)
%\end{equation}

\subsection{Graph Construction}
We construct a heterogeneous graph $\mathcal{G}$ to hold the semantic relatedness between different entities in Keyword-RS. Specifically, there are three types of nodes: user, paper, and keyword. We split the title of the paper into individual words\footnote{Here we use words in the title rather than keywords, the reason is that there are obstacles in obtaining the keyword information, which is absent in a part of papers.} and filter out stopwords by word tokenization tool like nltk\footnote{https://github.com/nltk/nltk}, and then gather these words as the keyword set $\mathcal{K}$. Then three edge types are built from raw interactions between nodes as follows:
 \begin{itemize}
 	\item \textbf{user-paper}. When user $u$ clicks or searches paper $p$, an edge $(u,p)$ is added.
 	\item \textbf{user-keyword}: When user $u$ adds keyword $k$ as shown in figure \ref{fig:Keyword-RS}, an edge $(u,k)$ is added.
 	\item \textbf{paper-keyword}: When the title of paper $p$ contains the keyword $k$,  an edge $(p,k)$ is added.
 \end{itemize}
In summary, $\mathcal{G}$ is defined as $\mathcal{G}=\{ \mathcal{V}, \mathcal{E}, O_{\mathcal{V}}, O_{\mathcal{E}} \}$ with nodes $\mathcal{V}$ and edges $\mathcal{E}$. And $O_{\mathcal{V}}=\{user, keyword,paper\}$ and $O_{\mathcal{E}}=\{u-p,u-k,p-k\}$ denote the set of node types and that of edge types. Note that all edges indicate a symmetric relation, which means $\mathcal{G}$ is an undirected graph. Figure \ref{fig:AMinerGNN}(a) shows an example of the above three situations. 
%figure? (a): construct graph (b): GNN (c): query attentive fusion (d): CTR

\section{METHODOLOGY}
\begin{figure*}[t]
	\centering
	\includegraphics[width=0.8\linewidth]{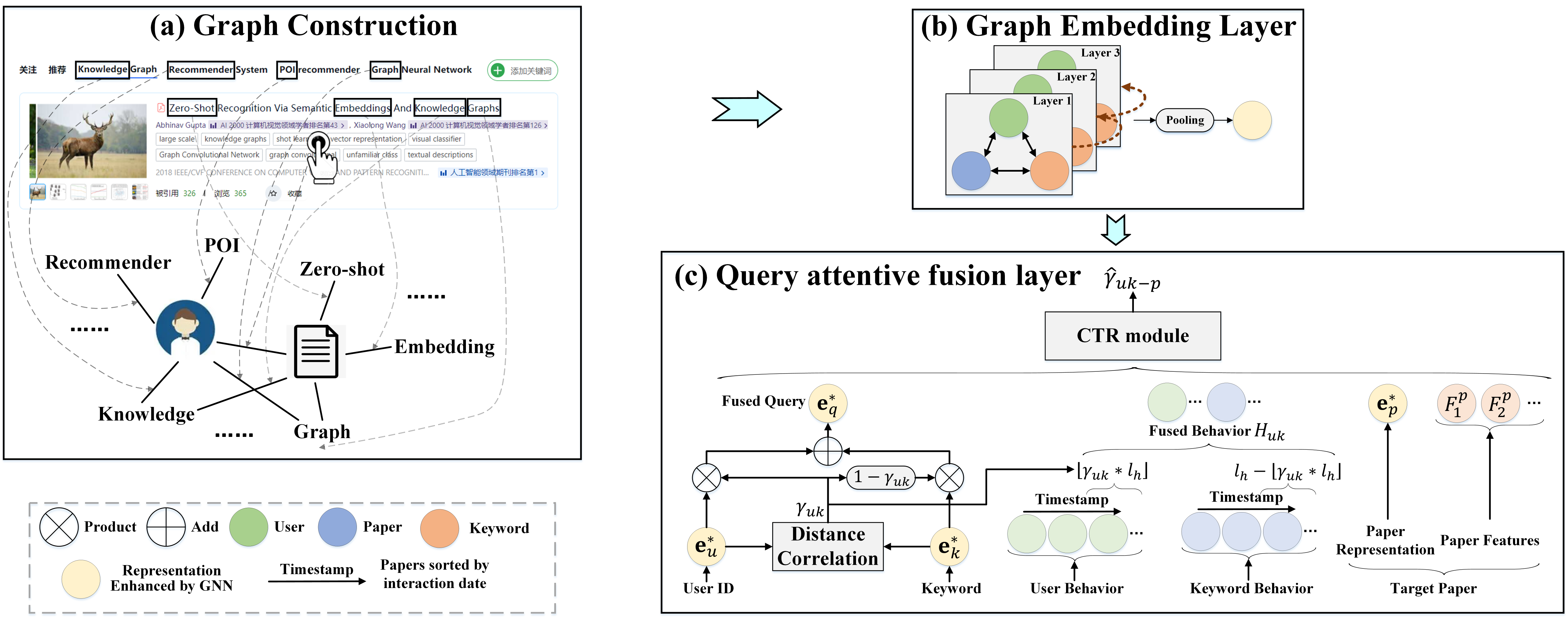}
	\caption{The framework of AMinerGNN. Best viewed in color.}
	\Description{The framework of AMinerGNN.}
	\label{fig:AMinerGNN}
\end{figure*}

In this section, we start from the motivations and intuitions, and then describe the technical details of the proposed AMinerGNN, which follows a hierarchical structure: GNN embedding layer $\rightarrow$ query attentive fusion layer. Figure \ref{fig:AMinerGNN} presents the framework of AMinerGNN. %Moreover, we exhibits the online service process of Keyword-RS.

\subsection{Motivation}
To tackle the two problems mentioned in Section \ref{sec:introduction}, a three-step solution is designed as follows: (1) representing user, paper, and keyword in the same embedding space $\rightarrow$ (2) enhancing representations via raw interactions between entities $\rightarrow$ (3) feeding user and keyword representations into a unified CTR model. 

The first two steps and last step are implemented in Graph embedding Layer (cf. Section \ref{sec:GEL}) and query attentive fusion layer (cf. Section \ref{sec:QAFL}), respectively.

Note that GNN is of great significance. Specifically, there are three advantages on different entity sides: (1) \textbf{On the paper side}. Paper representations, which keywords are integrated into, can contain the information of the research direction of the paper. (2) \textbf{On the user side}. The interacted papers of a user represent his past research direction and preferences in viewing papers. What's more, the keywords user adds directly reflect his concerned research field. Therefore, it is pivotal to enhance user representations with the above two entities. (3) \textbf{On the keyword side}. If two keywords co-occur in one paper, it shows the potential of interdisciplinary research between them. Thus, mutually enhanced representations of a pair of keywords can bring them closer in latent semantic space.  The closer the distance between two keyword embeddings, the greater the potential of interdisciplinary research.

\subsection{Graph  embedding Layer} \label{sec:GEL}
To achieve the first two steps, GNN is an intuitive tool to enhance node representation using multi-hop neighbors' information. In each layer, we employ a widely used two-step scheme\cite{wang2019heterogeneous,zhang2019heterogeneous,fu2020magnn} to aggregate neighbors: (1) same relation-aware neighbors aggregation via average pooling; (2) relations combination via sum pooling. And we leave the further work of other pooling methods like attention or metapath based aggregation as the future work. More formally, in the $l$-th layer, node representation is updated as
\begin{equation}
	\label{eq:atr_pro_1}
	\begin{aligned}
		\mathbf{e}^{(l)}_i=\sum_{t \in \mathcal{O_{\mathcal{E}}}} \frac{1}{\mid \mathcal{N}_i^{t} \mid} \sum_{j \in \mathcal{N}_i^{t} } \mathbf{e}^{(l-1)}_j
	\end{aligned}
\end{equation}
where $\mathbf{e}^{(l)}_i$ denotes the embedding of node $i$ in the $l$-th layer and $\mathbf{e}^{(0)}_i$ is randomly initialized. Here $i$ can be one of the user, paper or keyword node. $\mathcal{N}_i^{t}$ denotes the $t$-type neighbors for node $i$ and $t \in \mathcal{O_{\mathcal{E}}}$.

After performing $L$ layers, we obtain multiple node representations and then adopt sum pooling to integrate multi-hop information as 
\begin{equation}
	\label{eq:KGAN0_f}
	\begin{aligned}
			\mathbf{e}^{*}_i=\sum_{l=0}^{L}\mathbf{e}^{(l)}_{i}
		\end{aligned}
\end{equation}
where superscript $*$ indicates that the representation is enhanced by GNN.

\subsection{Query Attentive Fusion Layer}\label{sec:QAFL}
With all representations in the same embedding space, we calculate the correlation between a user and a keyword, denoted as $\gamma_{uk}$, using distance correlation as
\begin{equation}
	\label{eq:discor}
	\begin{aligned}
		\gamma_{uk}=d \operatorname{Cor}\left(\mathbf{e}^{*}_{u}, \mathbf{e}^{*}_{k}\right)=\frac{d \operatorname{Cov}\left(\mathbf{e}^{*}_{u}, \mathbf{e}^{*}_{k}\right)}{\sqrt{d \operatorname{Cov}\left(\mathbf{e}^{*}_{u},\mathbf{e}^{*}_{u} \right) \cdot d \operatorname{Cov}\left(\mathbf{e}^{*}_{k},\mathbf{e}^{*}_{k}\right)}}
	\end{aligned}
\end{equation}
where $d \operatorname{Cov}(\cdot)$ is the distance covariance of two representations.

As aforementioned in Section \ref{sec:introduction}, the relation between two queries is non-uniform for all scenarios. More specifically, \textbf{the importance of user ID is in proportion to the correlation between two queries}. Therefore, an intuitive idea is regarding correlation $\gamma_{uk}$ as the attention weight to form a unified and fused query as
\begin{equation}
	\label{eq:query}
	\begin{aligned}
		\mathbf{e}^{*}_{q}=\gamma_{uk} \mathbf{e}^{*}_{u}+(1-\gamma_{uk}) \mathbf{e}^{*}_{k}
	\end{aligned}
\end{equation}

Note that $\mathbf{e}^{*}_{q}$ can represent user purposes in different scenarios. For example, consider the limit case, when $\gamma_{uk}=0$ which means the user-generated keyword is absolutely unrelated to his past research direction, $\mathbf{e}^{*}_{q}=\mathbf{e}^{*}_{k}$ is reasonable because there is no need to use personal interest for recommendation at this situation. It has similar rationality in the other two scenarios.

After two queries are fused, another problem is how to process query behaviors $H_u$ and $H_k$. Similar to the fusion of $\mathbf{e}^{*}_{u}$ and $\mathbf{e}^{*}_{k}$, a heuristic idea is to combine a partial sequence of each behavior as one new behavior, denoted as $H_{uk}$. The length of each partial behavior is controlled by $\gamma_{uk}$. For example, when $\gamma_{uk}$ is large, we hope to reserve more user behaviors and less keyword behaviors, and vice versa. For faster and unified tensor calculation, we set the length of $H_{uk}$ as a fixed number $l_h$.  Then $\lfloor\gamma_{uk}*l_h\rfloor$ recent papers in $H_u$ and $l_h-\lfloor\gamma_{uk}*l_h\rfloor$ recent papers in $H_k$ are selected to form $H_{uk}$. The notation $\lfloor x \rfloor$ refers to the greatest integer that is less than or equal to $x$. Formally, $H_{uk}$ is generated as
\begin{equation}
	\label{eq:behavior}
	\begin{aligned}
		H_{uk}=\{p_1,p_2|p_1 \in H_u, p_2 \in H_k, |p_1|=\lfloor\gamma_{uk}*l_h\rfloor=l_h-|p_2| \}
	\end{aligned}
\end{equation}

Finally, we fed fused query $\mathbf{e}^{*}_{q}$ and bahavior $H_{uk}$, and other auxiliary information $F^{p}$ into a CTR module to predict the probability of paper $p$ being clicked as
\begin{equation}
	\hat{y}_{uk-p}=f_{\operatorname{CTR}}(\mathbf{e}^{*}_{q},\mathbf{e}^{*}_{p},H_{uk},F_p)
\end{equation}

where $f_{\operatorname{CTR}}$ can be any CTR model, which is set as DIEN \cite{zhou2019deep} by default.

Note that AMinerGNN can be generalized to CTR prediction with multi queries. For example, when the user inputs two keywords, the idea of query fusion is still productive to combine three queries (one user ID and two keywords) as one attentively, and three attention weights are distributed according to their importances, which we leave as future work.

\subsection{Model Optimization}
We use Logloss, which is the most used objective function in CTR prediction, to capture divergence between two probability distributions as
\begin{equation}
	\label{eq:loss}
	\begin{aligned}
		\mathcal{L}=-\frac{1}{N} \sum_{(u,k,p) \in S}(y_{uk-p}\log{\hat{y}_{uk-p}}+(1-y_{uk-p})\log(1-\hat{y}_{uk-p}))
	\end{aligned}
\end{equation}
where $S$ is the training set of size $N$.

\section{experiments}
\subsection{Settings}
\subsubsection{Dataset Description} We propose a novel dataset, called AMiner\_KRS dataset. We collect the interaction data in the past three months, from Feb. 1, 2022, to May 1, 2022. To ensure data quality, we filter the unregistered users and low-quality papers\footnote{AMiner has collected about billions of papers stored in MongoDB database. However, most of them are viewed by researchers on our website less than 5 times. Therefore, we just retain the high-quality papers by setting the threshold of citation number and view number.}. Finally, AMiner\_KRS consists of 67806 users, 398001 papers, and 200837 keywords.To avoid data leakage, we use the former two months' data to construct the graph and generate query behaviors, while the user\&keyword-paper interaction data in the third month (Apr. 2022) is randomly split into training (80\%) and testing (20\%) sets.  Table \ref{tab:Statistics of the AMiner_KRS dataset} lists the statistics of the AMiner\_KRS dataset.
\begin{table}[]
	\caption{Statistics of the AMiner$\_$KRS dataset. 'u', 'k' and 'p' denote user, keyword and paper, respectively. }
	\label{tab:Statistics of the AMiner_KRS dataset}
	\begin{tabular}{ccccc}
		\hline
		%		1&1&1&1&1
		\multicolumn{1}{c}{\multirow{4}{*}{Graph $\mathcal{G}$}}  & \multicolumn{1}{|c|}{\multirow{2}{*}{node}}  & \multicolumn{1}{c}{\# users}& \multicolumn{1}{c}{\# keywords} & \multicolumn{1}{c}{\# papers} \\
		& \multicolumn{1}{|c|}{} & \multicolumn{1}{c}{67,806} & \multicolumn{1}{c}{398,001} & \multicolumn{1}{c}{200,837} \\
		\cline{2-5}
		& \multicolumn{1}{|c|}{\multirow{2}{*}{edge}}  & \multicolumn{1}{c}{\# u-p}& \multicolumn{1}{c}{\# u-k} & \multicolumn{1}{c}{\# p-k}\\
		& \multicolumn{1}{|c|}{} & \multicolumn{1}{c}{568,424} & \multicolumn{1}{c}{240,116} & \multicolumn{1}{c}{6,899,284} \\
		\hline\hline 
		\multicolumn{2}{c|}{\multirow{2}{*}{Interactions $\mathcal{Y}$}} & \multicolumn{1}{c}{\# $y_{u-p}$}  & \multicolumn{1}{c}{\# $y_{k-p}$}& \multicolumn{1}{c}{\# $y_{uk-p}$}\\
		&\multicolumn{1}{c|}{} & \multicolumn{1}{c}{568,424} & \multicolumn{1}{c}{5,030} & \multicolumn{1}{c}{2,781} \\
		\hline           
	\end{tabular}
\end{table}

\subsubsection{Baselines} 
To demonstrate the effectiveness, we compare AminerGNN with two kinds of methods: (1) \textbf{CTR} models including DeepFM \cite{guo2018deepfm}, DIN \cite{zhou2018deep}, DIEN \cite{zhou2019deep}, UBR \cite{qin2020user}), (2) \textbf{Product search} methods including HEM \cite{ai2017learning}, ALSTP  \cite{guo2019attentive}, MGDSPR \cite{li2021embedding}.

\subsubsection{Experimental Settings} 
The core contribution of AMinerGNN is generalizing traditional CTR models to the task with fusion query using two key designs: (1) GNN embedding layer and (2) query fusion layer. Therefore, our experiments aim to prove the effectiveness of these two parts. Specifically, for base model, two queries are added to one embedding and two behaviors are fused into an union set, which are directly fed into CTR model, such as $f_{\rm DIN} (\mathbf{e}_{u} + \mathbf{e}_{k},\mathbf{e}_{p},H_u \cup H_k,F_p)$. Then we compare each model with the following three variants: (1) using graph embedding layer, (2) using query attentive fusion layer, and (3) both adopting these two layers. We use subscript "base", "g", "f" and "g$\&$f" to denote the base model and corresponding three kinds of variants. Note that AMinerGNN is identical to $\rm DIEN_{g\&f}$. 

The schema of input in product search methods usually has three sequences, which are users, corresponding queries, and corresponding products. We use $y_{uk-p}$ to generate the above three sets. For ALSTP, we search the number of previous queries during short-term preference modeling in the range of $\{16,32,64,100\}$. For MGDSPR, the window sizes of long and short-term sequences are ten days and one month, respectively. 

For all models, the embedding size is fixed to 64 and $l_h$ is set to be 100. We optimize our method with Adam \cite{kingma2014adam} and use the default learning rate of 0.001 and default mini-batch size of 1024. The number of GNN layer $L$ is set to be 2 for the best performance.%\footnote{We search $L$ in $\{1, 2, 3\}$ and $L=2$ leads to the best performance.}.

\subsubsection{Metrics} For CTR prediction task, we adopt the commonly used AUC and Log Loss to measure the performance.

\subsection{Performance Comparison}
As shown in Table \ref{tab:Overall comparision},  We have the following observations:
\begin{itemize}
	\item For CTR baselines, there are similar performances and we use DIEN as an example to illustrate our findings. AMinerGNN and $\rm DIEN_{f}$ contribute up to 5.02\% AUC and 4.11\% AUC promotion compared to $\rm DIEN_{g}$ and $\rm DIEN$, which proves the effectiveness of the query fusion layer. AMinerGNN and $\rm DIEN_{g}$ improves over $\rm DIEN_{f}$ and $\rm DIEN$ w.r.t. AUC by 26.44\% and 23.24\%, which proves the effectiveness of GNN embedding layer. Moreover, we find that the improvement brought by the GNN embedding layer is more significant than that by the query fusion layer. The reason is that the GNN embedding layer projects all entities into the same embedding space, which is the prerequisite to accurately modeling the semantic relation between two queries using distance covariance. Another possible reason is that the number of whole keyword behaviors is particularly less than that of user behaviors (in Table \ref{tab:Statistics of the AMiner_KRS dataset}). This problem may be improved by the increase of Aminer daily UV (unique visitor) in the future. 
	\item  For product search baselines, all of them underperform than $\rm DIEN_{g\&f}$ and $\rm UBR_{g\&f}$. The main reason is that they insufficiently recognize the pattern that the user attention towards the input keyword and personalization is non-uniform in different scenarios. A detailed comparison is reported in Section \ref{sec:st}.
	\item For two kinds of baselines, modeling temporal interests brings in great improvements, like DIEN compared to DIN, and ALSTP compared to HEM. A clear reason is that the research direction of one user is changing with time.
	\item Note that we adopt DIEN as our base CTR model rather than UBR. $\rm UBR_{g\&f}$ slightly outperforms $\rm DIEN_{g\&f}$. However, the online service time of $\rm UBR_{g\&f}$ is much bigger than that of $\rm DIEN_{g\&f}$, because UBR uses the search engine approach to retrieve the user behaviors, leading to an increase of the online inferring time. We use Locust\footnote{https://locust.io/} to simulate 200 simultaneous users and average the response time of CTR-based models\footnote{We use TorchServe (https://github.com/pytorch/serve) to deploy them.}, which are shown in Table \ref{tab:Online time}. Finally, $\rm DIEN_{g\&f}$ is deployed in AMiner as a trade-off between performance and speed.
\end{itemize}

\begin{table}[]
	\caption{Overall comparison.}
	\label{tab:Overall comparision}
	\begin{tabular}{c|cc}
		\hline
		\multicolumn{1}{c|}{}  & AUC & Log Loss \\
		\hline\hline
		\multicolumn{1}{c|}{$\rm DeepFM_{base/f/g/g\&f}$} & 0.635/673/799/837 & 0.674/666/566/517 \\
		\multicolumn{1}{c|}{$\rm DIN_{base/f/g/g\&f}$} &0.657/684/856/876 & 0.645/642/496/465 \\
		\multicolumn{1}{c|}{$\rm DIEN_{base/f/g/g\&f}$} & 0.710/711/875/899 & 0.581/577/470/452\\
		\multicolumn{1}{c|}{$\rm UBR_{base/f/g/g\&f}$} & 0.716/719/876/903 & 0.578/571/473/450\\
		\hline\hline
		\multicolumn{1}{c|}{HEM} & 0.672 & 0.680 \\
		\multicolumn{1}{c|}{ALSTP} & 0.840 & 0.512 \\
		\multicolumn{1}{c|}{MGDSPR} & 0.871 & 0.478 \\
		\hline\hline
		\multicolumn{1}{c|}{AMinerGNN(a.k.a. $\rm DIEN_{g\&f}$)} & \textbf{0.899} & \textbf{0.452} \\
		\hline           
	\end{tabular}
\end{table}

\begin{table}[]
	\caption{Online response time.}
	\label{tab:Online time}
	\begin{tabular}{c|ccc}
		\hline
		\multicolumn{1}{c|}{}  & $\rm DIN_{g\&f}$& $\rm DIEN_{g\&f}$ & $\rm UBR_{g\&f}$ \\
		\hline\hline
		\multicolumn{1}{c|}{time(ms)} & 97 & 106 & 329 \\
		\hline           
	\end{tabular}
\end{table}

\subsection{Scenario Test} \label{sec:st}
\begin{figure}[t]
	\centering
	\includegraphics[width=0.7\linewidth]{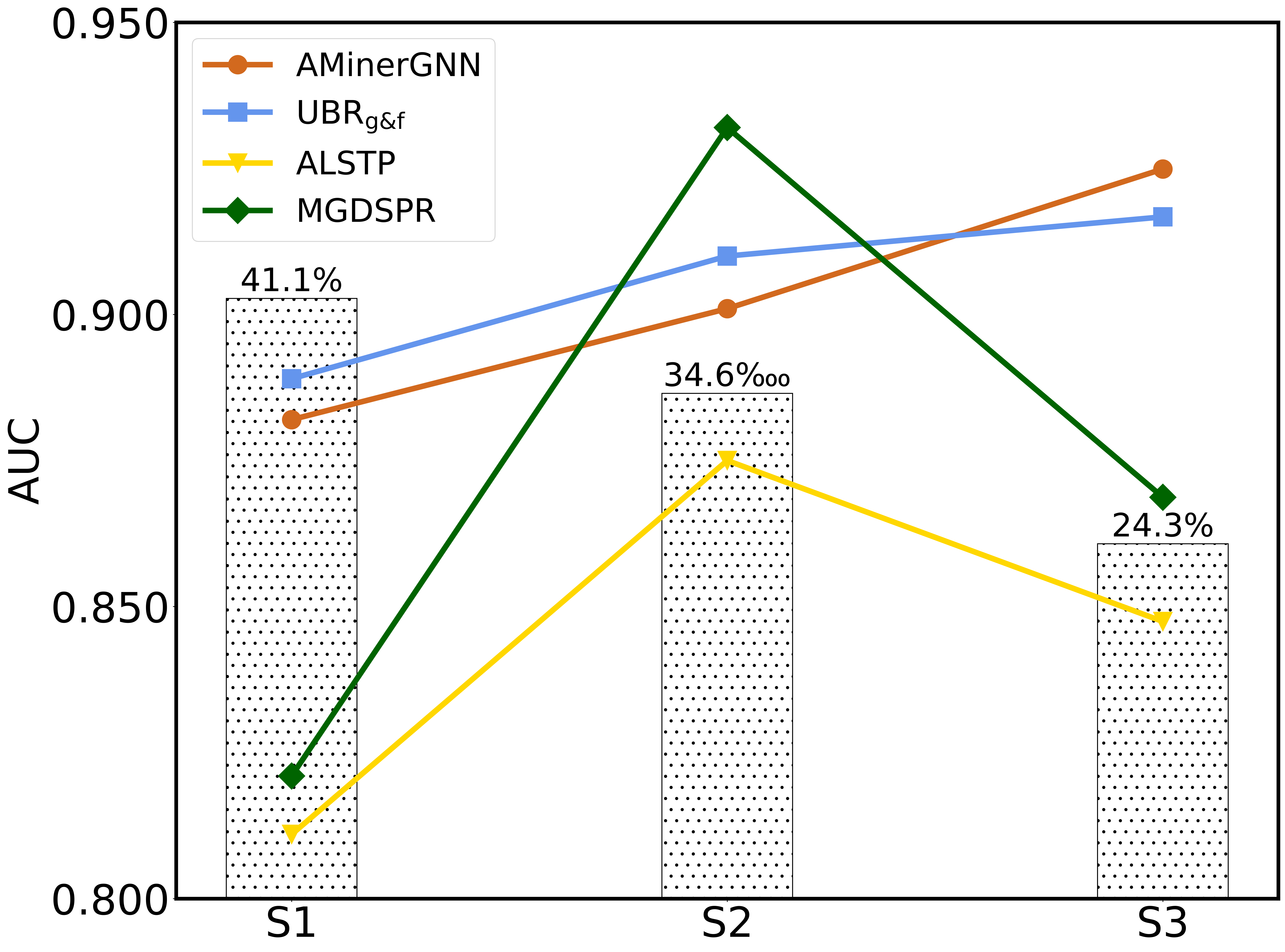}
	\caption{Scenario test. The background histogram represents the proportion of the number of the testing batch in each scenario.}
	\Description{The framework of AMinerGNN.}
	\label{fig:stest}
\end{figure}

In this section, we investigate whether AminerGNN adaptively fits for different scenarios. We calculate $\gamma_{uk}$ via equation \ref{eq:discor} for each testing batch. $\gamma_{uk} \in [0.5,1], [0.1,0.5),[0,0.1)$ indicate \textbf{S1}, \textbf{S2}, and \textbf{S3}, respectively. The results are shown in Figure \ref{fig:stest} and the major findings are as below:
\begin{itemize}
	\item CTR-based methods perform stably in three scenarios, which proves that two key components of AminerGNN can generalize traditional CTR models to a unified recommender system applicable in all scenarios.
	\item Product search-based models perform better in  \textbf{S2}, while the performances in the other two scenarios sharply decrease. The reason is the input keyword and personalization are both important in  \textbf{S2}, which can realize their potentials best. While in the other two scenarios, only one query dominates influence, which is less applicable for product search-based models.
\end{itemize}

\subsection{Online A/B Test}
We report the online A/B tests of AMinerGNN for a duration of 14 days, from May 17, 2022, to May 30, 2022. About 30 thousand users participated in this A/B test. Two kinds of click-through rate (CTR) on the homepage of AMiner are chosen as metrics. ‘UV-CTR’ indicates whether a user clicks the recommended paper, irrespective of the number of clicks. 'PV-CTR' is the ratio of clicked papers to all exposed papers. As in Table \ref{tab:Online CTR}, AMinerGNN achieves 9.6\% and 3.7\% UV-CTR improvement relative to DIEN and MGDSPR, which demonstrates AMinerGNN works better in paper recommendation with the keyword.

\begin{table}[]
	\caption{Online CTR of compared methods.}
	\label{tab:Online CTR}
	\begin{tabular}{c|ccc}
		\hline
		\multicolumn{1}{c|}{Metrics}  & DIEN & MGDSPR& AMinerGNN \\
		\hline\hline
		\multicolumn{1}{c|}{UV-CTR} & 0.437 & 0.496 & 0.533 \\
		\multicolumn{1}{c|}{PV-CTR} & 0.086 & 0.097& 0.112 \\
		\hline           
	\end{tabular}
\end{table}
% We report the confidence intervals under the numbers using students’ t-test with p-value far less than 0.005.

\section{conclusion}
In this paper, we study the recommendation task with two queries. The proposed AMinerGNN can automatically mine user scenario-specific purposes, dynamically recognize the importances of two queries, and attentively fuse two queries to construct a unified and end-to-end recommender system. Compared with the traditional CTR model given one query and product search methods, AMinerGNN performs better and achieves a higher CTR on the AMiner homepage.

\bibliographystyle{ACM-Reference-Format}
\bibliography{ref1}

\end{sloppypar}
\end{document}